\documentclass[11pt,a4paper,headinclude,footinclude,chapterprefix=on]{scrreprt}
\usepackage{graphicx}
\usepackage[utf8]{inputenc}

\usepackage[procnames]{listings}
\usepackage{color}

\usepackage{amssymb}
\graphicspath{{figures/}}

\usepackage{url}
\usepackage{wrapfig}
\usepackage{units}

\usepackage[printonlyused]{acronym}

\usepackage{calc}
\usepackage{amsmath}
\usepackage{amssymb}
\usepackage{amsfonts}
\usepackage{algorithm}
\usepackage{algpseudocode}
\usepackage{color}
\usepackage[table]{xcolor} %
\usepackage{caption}
\usepackage{subcaption}
\usepackage{stfloats}
\usepackage{nicefrac}

\usepackage{booktabs}
\newcolumntype{C}[1]{>{\centering\arraybackslash}p{#1}}

\usepackage{color}

\newcolumntype{C}[1]{>{\centering\arraybackslash}p{#1}}

\usepackage{booktabs}

\definecolor{BgGray}{gray}{0.7}%
\definecolor{BgGray2}{gray}{0.96}%
\definecolor{RowColorOdd}{named}{BgGray2}%
\definecolor{RowColorEven}{named}{white}%
\definecolor{comments}{gray}{.5}
\definecolor{Gray}{gray}{0.85}

\definecolor{keywords}{RGB}{255,0,90}
\definecolor{red}{RGB}{160,0,0}
\definecolor{green}{RGB}{0,150,0}
\definecolor{deepblue}{rgb}{0,0,0.5}
\definecolor{deepred}{rgb}{0.6,0,0}
\definecolor{deepgreen}{rgb}{0,0.5,0}

\usepackage{multirow}

\usepackage[utf8]{inputenc}
\usepackage{tabularx}
\usepackage{multicol}
\usepackage{graphicx}

\usepackage{tikz}

\usepackage[headsepline,plainheadsepline,footsepline,plainfootsepline]{scrpage2}
\usepackage{mathptmx}
\usepackage[scaled=.92]{helvet}
\usepackage{courier}
\usepackage[multiuser,nomargin,marginclue,footnote]{fixme}

\definecolor{BgGray}{gray}{0.1}%
\definecolor{BgGray2}{gray}{0.96}%
\definecolor{RowColorOdd}{named}{BgGray2}%
\definecolor{RowColorEven}{named}{white}%
\definecolor{comments}{gray}{.5}
\definecolor{Gray}{gray}{0.85}

\usepackage{pifont}%
\usepackage[multiuser,nomargin,marginclue,footnote]{fixme}
\fxusetheme{color}
\fxsetup{targetlayout=colorcb}
\FXRegisterAuthor{all}{anall}{ALL}
\FXRegisterAuthor{tolja}{antolja}{TOLJA}
\FXRegisterAuthor{mch}{anmc}{MC}
\FXRegisterAuthor{cp}{ancp}{CP}

\fxsetup{status=draft}

\definecolor{BgGray}{gray}{0.5}%

\usepackage{geometry}
\clearscrheadfoot

\ihead[\Large {\scshape TU Berlin }]{\Large {\scshape TU Berlin }}

\ifoot[{\tiny \begin{minipage}{4.0cm}Copyright at Technische Universität Berlin.\newline All Rights Reserved.\end{minipage}}]{{\tiny \begin{minipage}{4.0cm}Copyright at Technische Universität Berlin.\newline All Rights Reserved.\end{minipage}}}
\ofoot[Page \pagemark]{Page \pagemark}

\addtokomafont{pagefoot}{\normalfont}

\newcommand{\trnumber}{TKN-15-0005}
\newcommand{\trdate}{Dezember 2015}
\newcommand{\trauthor}{Sven Zehl, Anatolij Zubow, Adam Wolisz and Michael Döring}
\newcommand{\tremail}{\{zehl, zubow, wolisz, doering\}@tkn.tu-berlin.de}
\newcommand{\trtitle}{ResFi: A Secure Framework for Self Organized Radio Resource Management in Residential WiFi Networks}

\begin{document}


{
\sffamily

\thispagestyle{empty}

\begin{tabularx}{\columnwidth}{cXc}
  \includegraphics[height=1cm]{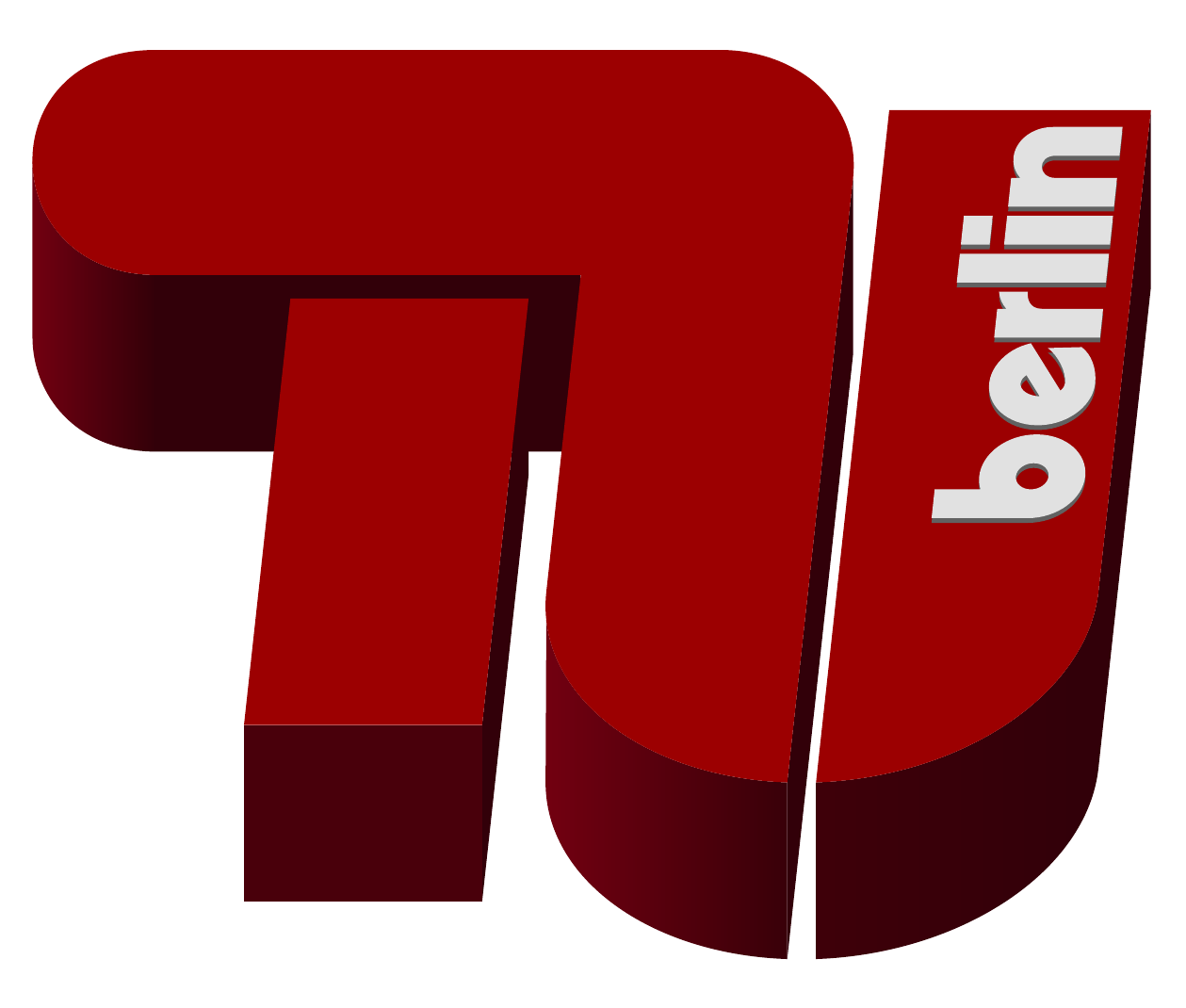}
  & &
  \includegraphics[height=1cm]{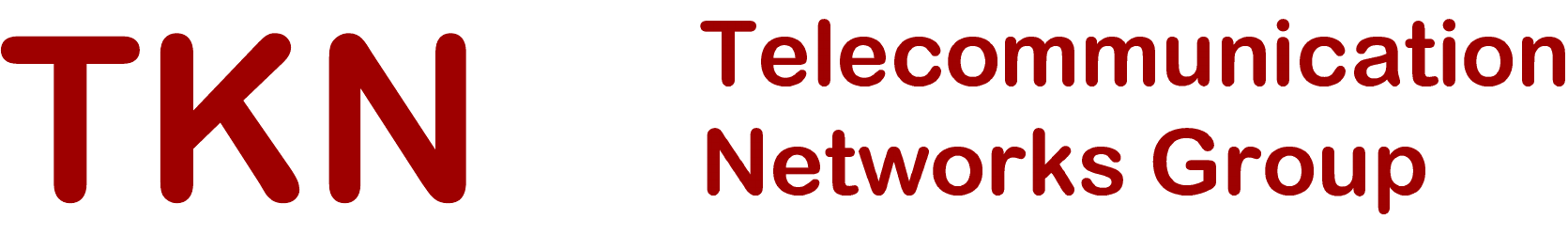}
  \\
\end{tabularx}

\vspace{1.0cm}

\begin{center}
{\huge
\noindent
Technische Universität Berlin

\vspace{0.5cm}

\noindent
Telecommunication Networks Group

\begin{center}
\rule{15.5cm}{0.4pt}
\end{center}
}
\end{center}

\begin{minipage}[][11.0cm][c]{14.5cm}
{\Huge

\begin{center}
\trtitle
\end{center}

\begin{center}
{\LARGE \trauthor} \\
{\Large \tremail}
\end{center}

\begin{center}
Berlin, \trdate
\end{center}

\vspace{0.5cm}

}

\begin{center}
\setlength{\fboxrule}{2pt}\setlength{\fboxsep}{2mm}
\fbox{TKN Technical Report \trnumber}
\end{center}

\end{minipage}

\setlength{\fboxrule}{0.4pt}
\setlength{\fboxsep}{0.4pt}

\begin{center}

  \rule{15.5cm}{0.4pt}

  \vspace{0.5cm}

  {\huge {TKN Technical Reports Series}}

  \vspace{0.5cm}

  {\huge Editor: Prof. Dr.-Ing. Adam Wolisz}

  \vspace{0.5cm}

 \end{center}

}

\begin{abstract}
\subsection*{\abstractname}
In dense deployments of residential WiFi networks individual users suffer performance degradation due to both contention and interference. While Radio Resource Management (RRM) is known to mitigate this effects its application in residential WiFi networks being by nature  unplanned and individually managed creates a big challenge.

We propose ResFi - a framework supporting creation of RRM functionality in legacy deployments. The radio interfaces are used for efficient discovery of adjacent APs and as a side-channel to establish a secure communication among the individual Access Point Management Applications within a neighborhood over the wired Internet backbone.

We have implemented a prototype of ResFi and studied its performance in our testbed. As a showcase we have implemented various RRM applications among others a distributed channel assignment algorithm using ResFi. ResFi is provided to the community as open source.

\end{abstract}

\tableofcontents

\chapter{Introduction}\label{sec:introduction}

In recent years we have seen a rapid growth in the use of wireless devices such as laptops, tablets and smart phones in all environments e.g., enterprise and homes. Especially, the IEEE 802.11 (WiFi) wireless technology gained lot of popularity as a comfortable way to connect a multitude of devices. As applications like mobile HD video \& cloud storage require high QoS, dense deployments of wireless technologies observed nowadays cause performance issues due to high contention and interference within the limited set of radio frequencies.
In enterprise networks remaining within a single administrative domain this issue is commonly solved by installing a centralized controller which manages the usage radio resources by all APs~\cite{murty2008designing,Zubow15bigap_seamless_handover}. 
The performance of this controller depends on the scope of information used – this is at least the sum of the traffic and channel usage observations by all the APs but gradually a trend to use also information provided by the end systems (using e.g. 802.11k) becomes also visible.  It has been widely demonstrated that the coordinated usage of radio resources has led to very significant improvement of the QoS, and in fact it is a fundamental condition to achieve satisfactory QoS in dense, heavily used environments. 
In contrast apartment house deployments usually consist of multiple autonomous APs remaining under administration of individual users. Indeed, each AP is usually installed by a resident who due to lack of technical skills attempts to minimize the configuration effort. While in the past this led to the well known phenomenon of using mostly the single, pre-set channel, manufacturers started increasing the scope of self-configuration functions provided. 
The scope of this self-configuration is, however, still limited to functions depending exclusively on local observations within this AP and local controls.  In the residential deployment the individual APs - even located in close proximity do not have a direct way to enter an organized information exchange and negotiations. In addition the usual consumer electronic devices expected in an apartment usually do not support management features like those provided by 802.11k, so that no additional information from them can be obtained.

In this paper we present ResFi - a set of basic self configuration functionalities enabling radio resource management in residential WiFi.
ResFi offers the following functionalities:
\begin{enumerate}
\item Discovery of the immediate neighborhood – any active APs within the radio coverage. 
\item Setting up secured point-to-point control channels between any pair of immediate neighbors over the wired Internet backbone.
\item Exchange of N-hop neighborhood information and continuous monitoring of the neighborhood using the above channels.  
\end{enumerate} 
ResFi is specified and implemented in form of platform independent source code which can be used on top of the legacy APs. Up to our best knowledge this is a first attempt to suggest such a platform. We believe that this set of basic function creates a good foundation to develop management application algorithms which itself is explicitly a NON goal of this paper.
As a proof of concept we provide however 
\begin{itemize}
\item A description, implementation and evaluation of a simple distributed AP channel assignment algorithm.
\item A description, implementation and evaluation of a simple distributed clustering algorithm to show how one could select a group of access point subjects to joint RRM.
\end{itemize} 
The performance of the proposed approach is evaluated by means of experiments in a real testbed. Moreover, we provide an emulation in Mininet~\cite{mininetUrl} which gives the developer an easy way to test own algorithms before  deploying  them  in  a  real  testbed.  Finally  ResFi  is  provided to the community as open source under GPL license on Github \url{https://github.com/resfi}.

\chapter{Related Work}\label{sec:related_work}

WiFi enterprise networks are already widely deployed in companies, universities and public spaces like airports and fairgrounds. 
Commercial enterprise WiFi solutions mostly feature a centralized controller which performs RRM for all
attached APs. For example the widespread CISCO solution~\cite{cisco-rrm} works as follows: each of the APs sends  periodically on all the radio channels "neighbor search"  messages including the Internet protocol address of their responsible
controller and the identifier of the group they belong to within this controller. Neighboring APs forward the received
"neighbor search" message including their own AP identifier to their responsible controller (frequently via the wired control connection). This enables the controller to build a hearing map, group APs in RRM groups or to elect a leader controller for the RRM process.
Distributed approaches are less frequent - e.g. Aerohive~\cite{aerohive-rrm} uses a classical distributed leader election algorithm for RF channel assignment. If a newly started AP discovers other APs on his RF channel, it advertises its neighbor count via the wireless channel while listening for the advertisment of the other APs. Finally the AP with the most neighbors wins the right to use the channel. All others switch to the next RF channel and the aforementioned procedure repeats.

Different options to optimize the RRM in enterprise WiFi networks have been addressed in research papers. Again the use of a central controller using wireless propagation data, 
collected from all deployed APs~\cite{murty2008designing,murty2008architecture,shrivastava2009centaur2,yiakoumis2014behop,Zubow15bigap_seamless_handover} dominate the field. 
The centralized view is then used to make global decisions in terms of e.g. channel assignment. In addition, more advanced approaches also provide the possibility of load balancing and handover operations~\cite{murty2008designing, yiakoumis2014behop,Zubow15bigap_seamless_handover} or transmit power and rate adaption control \cite{murty2008architecture}.

A typical residential WiFi deployment usually consists of
statically deployed APs and mobile client STAs. As the APs are not administered by a single authority but rather
as each AP is independently managed by another unexperienced
user, residential WiFi deployments can be assumed as
chaotic~\cite{Akella-2005}. The density of APs is highly correlated with the residential density and large-scale measurements~\cite{biswas2015large} showed
that the number of neighboring APs is relatively high in
urban environments, i.e. each AP has on average around 16.8
neighboring APs in the 2.4 GHz band.
In this chaotic deployment there does NOT exist a natural way to establish the information exchange between each of them. There does not exist a dedicated controller, and the skills of the “human administrator” are usually limited.
Patro et al.~\cite{Patro-2015} have postulated the use of a  cloud-based controller for channel assignment and airtime management. They propose to run one controller per building either funded by an Internet service provider (ISP) or the building manager. Further, the interaction between the residential APs and the controller is enabled by an extended version of the OpenFlow protocol. This approach seems promising for single administered apartment houses (single ISP or single building manager) but due to the lack of an auto configuration possibility it has its difficulties for all other deployment scenarios. Besides, the funding of the centralized controller and the payment of its operational costs is not easy to clarify. A controller-less solution would be favorable. 

In RxIP~\cite{manweiler-2012} a novel approach: direct communication between neighboring APs is introduced for the first time. Each home AP transmits a globally-routable IP address through additional information embedded within the periodically broadcasted beacon frames. This allows passively listening neighbor APs to communicate with the transmitter over the wired Internet, thus featuring a P2P fashion of interaction. RxIP does not aim RRM in general but rather targets the specific use-case of hidden terminal discovery and mitigation of its effect. In dialog with its neighbors, each AP collects independently the information about potential hidden terminals related to him. Therefore the RxIP approach is by definition restricted to discovery of only those neighboring APs which use the 
same RF channel. Nevertheless a more global view seems to be desirable.
Using large-scale measurement data from several cities Akella et al.~\cite{Akella-2005} showed that end-client experience in home WiFi networks could be significantly improved by managing the transmit power in such chaotic wireless networks. Using their proposed load-sensitive rate fallback implementation (LPERF) in which transmitters reduce their transmit power even if it reduces their transmission rate, they were able to show significant throughput enhancement through interference reduction among neighboring APs in dense deployments by incorporating among others the traffic demands of neighboring APs. They did, however not provide suggestions how the relevant stations are to be selected and how should they exchange the necessary coordination information. 
Finally numerous papers have addressed distributed radio resource management. For example in~\cite{Kubisch-2003}
power assignment in arbitrary wireless topologies has been assigned in a distributed way. 
Nevertheless none of these papers investigates specifically HOW to assure connectivity needed for information exchange among the involved nodes.
\chapter{IEEE 802.11 Primer}\label{sec:dot11primer}

This section gives a brief overview of the relevant aspects of the IEEE 802.11 standard.

\section{AP Discovery (i.e. scanning)}
Two approaches are possible. 

\subsection{Passive Scanning}
All IEEE 802.11 APs are broadcasting beacon frames in a fixed time interval to announce the existence of the 802.11 network. This includes advertising the Service Set Identifier (SSID) as well as all parameters needed for STAs to identify whether a connection to the network is possible. Upon activation/arrival, client STAs move to each available RF channel, listen for beacon frames, buffer the embedded information and create a map of available APs (obviously an AP can passively scan the environment in the same way). Passive scanning requires spending quite a significant time while listening to each channel! Due to security consideration APs may, however, suppress the announcement of the SSID, therefore passive scanning does NOT assure discovery of ALL available networks.
\subsection{Active Scanning}\label{sec:probe}
STAs searching for 802.11 networks can send out \textit{probe request} frames. If an AP receives a probe request with a matching SSID and suggested data rates within the scope supported by this AP, it replies with a probe response including all parameters of the normal beacon frame. Interestingly, an STA (or AP interested in his neighborhood) can use in the request frame a broadcast SSID which triggers an response from all networks which have been able to receive this request.

\ %

\section{Information Elements}\label{sec:ie}

Beacon, probe request and probe response frames are defined as management frames. The frame body of a management frame is built up of fixed length fields and variable length fields which are called \textit{information elements (IE)}. %
In addition to the standard IE(s) which are used e.g. to transport the SSID or the supported data rates, the standard describes a specific one called \textit{vendor specific information} and allowing to transport up to 255 bytes of custom data. We will denote this specific IE(s) as IEV.

\chapter{ResFi Design Principles}\label{sec:bigap_design}

\section{System Model}

 \begin{figure}
 \centering
 \begin{minipage}[b]{0.75\linewidth}
    \begin{center}
        \includegraphics[width=0.7\linewidth]{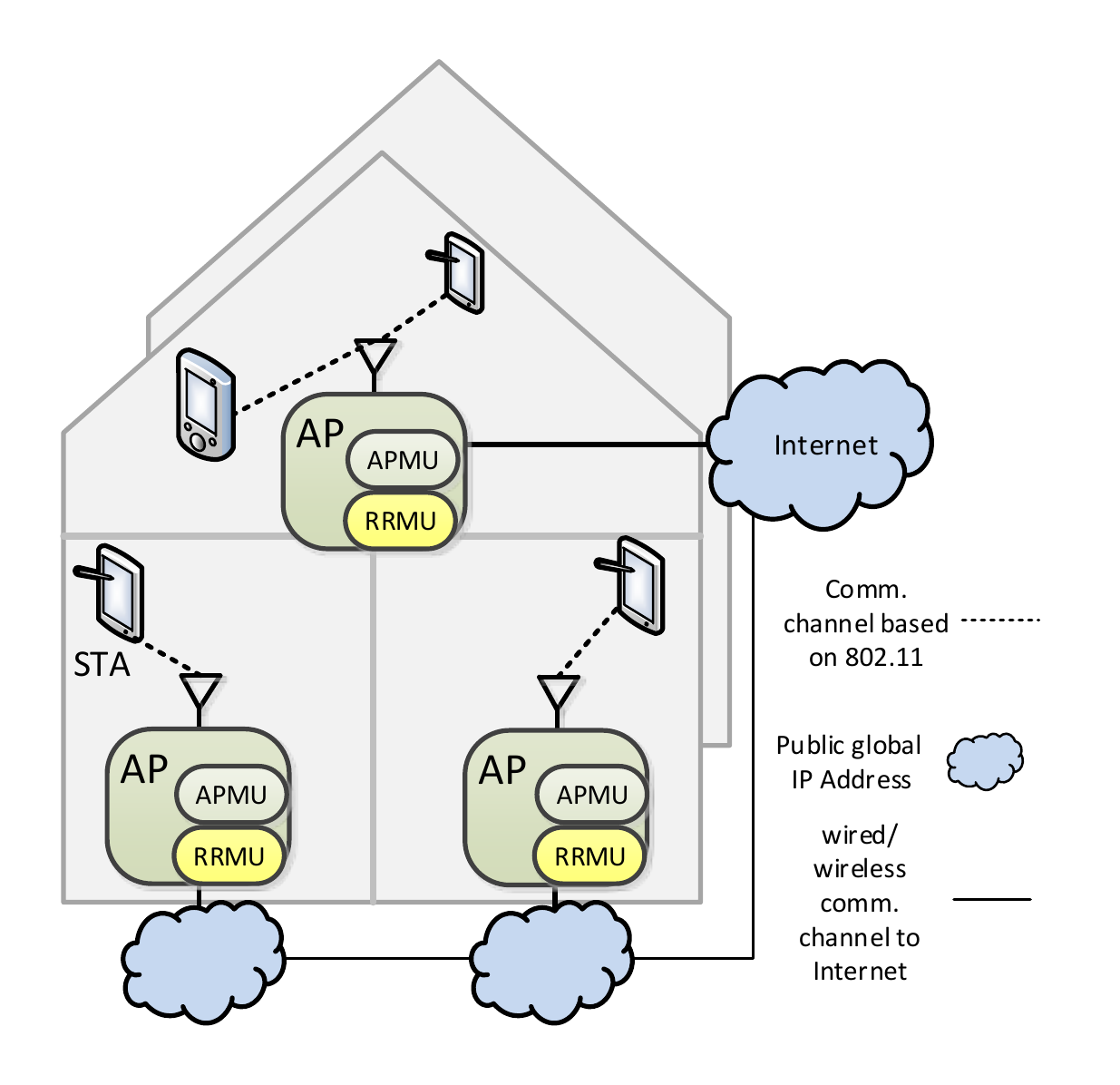}
    \end{center}
     \vspace{-15pt}
    \caption{A residential WiFi network consists of client Stations (STA) and Access Points (AP). Each AP is connected via wired broadband access to the Internet.}
    \label{fig:sys_model}
 \end{minipage}
 \vspace{-10pt}
 \end{figure}

Our view of the “chaotically deployed” WiFis is presented in Fig.~\ref{fig:sys_model}.

Each static AP is assumed to have two network interfaces, namely, an WiFi compliant air interface used for wireless communication towards STAs and a wired broadband access interface with a globally-routable IP address to connect to the Internet usually via DSL or cable modem. As we assume DSL/cable as the wired access technology there is a significant last-mile latency on the wired link to the first hop inside the ISP’s network~\cite{sundaresan2011broadband}.

We assume that the APs in a given neighborhood are deployed gradually (meaning they are switched on for the first time one by one), Any AP might also be switched off at any time – temporarily or for good.

We assume also that every AP is controlled by a AP management unit (APMU) which consists of several functional blocks such as client access control and operational parameter setting (like beacon interval setting). In classical deployment each AP has some - rather simplistic - local radio resource management e.g. setting of a fixed transmission channel, or simple selection of the transmission channel.
 
We postulate introducing in each AP a dedicated process called RRMU which is assumed to have IP connectivity over the wired Internet backhaul. Moreover, the RRMU is assumed to have an API (called southbound API) making it possible to access radio statistics and parameters within the AP. 

\section{Principles of the ResFi Framework}
The goal of ResFi is to define a self organized creation of a secured connectivity among the RRMU of all APs within a given neighborhood without: 
\begin{itemize}
\item Violating the assumptions of keeping each of the 
participating APs under separate local management
\item Any changes in the hardware and drivers of commercially available access points
\end{itemize}
The approach can be presented in a nutshell as follows: During the boot-up phase of any AP a broadcast scan request 
including a ResFi specific IEV containing so called "contact data" is triggered sequentially on each of the supported channels. Any AP within the coverage of this scan request is expected to answer with the respective "contact data" of the responder. These contact data, embedded in the IEV of both the active scan probe and response consists of the globally-routeable IP address and port number of the AP's RRMU (on the fixed internet) as well as of a transient 1-hop group encryption key and a public cryptography key individual to this RRMU.

After having completed the scan and having received the answers, the RRMU of the newly booted AP can establish a secure, point-to-point control channel to the RRMUs of all the "discovered" APs over the wired backbone Internet. Placing the control channel into the wired connectivity has several advantages. Notably there is no additional load on the wireless interfaces, and there is obviously a lower error rate. On the other hand longer message exchange delays have to be taken into account. This does not seem to be really a big issue, as the radio resource management does not take place in very short time scales.

Thus a coordination within one-hop neighborhood would be available at this point. 
It is, however, well known that RRM (e.g. channel selection) can achieve better efficiency if performed over a cluster of APs larger than one hop neighborhood. Therefore ResFi requires that each RRMU is able to act as a forwarder enabling to extend secure connectivity towards up to N hops (N can be set individually for every message sent via ResFi's northbound framework API). ResFi does not define the precise policy to create an RRM cluster within the scope of the connectivity borders mentioned above, neither does it feature a specific RRM approach. Both of these decisions are delegated to an RRM APPLICATION  which is not a part of the platform itself. We will provide in Section \ref{sec:resfi_applications} some examples of such applications. 

The security of the control channel is not constrained to the establishment with the use of proper cryptographic keys, in addition the keys are occasionally exchanged (see the following part).

\section{ResFi Security Model}

Why do we care about the security of the control channel for cooperative RRM? The reason is very simple. Divisive action might severely harm the wireless access of some users, and lead to an unfair advantage of some AP owners. While "unfair cheating" behaviors can not be completely eliminated (e.g some AP might claim that there are numerous APs in his vicinity thus luring neighbors to leave a channel for him alone) we offer within our framework a set of measures leading to clear identification of the source and destination of any information as well assuring the integrity of any information exchanged via the control channel. 
By the set of this means we can at least be sure, that the possible malicious behavior of any of the participants might be - after detection - uniquely traced back to this participant. And there will be no way this participant might claim his innocence. We will discuss below the threats we are considering - i.e the security model - adding a "rough outline" of the countermeasures. 

The primary exchange of security material for establishing a secure control channel takes place over the wireless channel within the exchange of the IEV in the probe request and probe reply frames. Therefore the possibility to get the security material is very constrained in space to the local observers.
 
\subsection{Thread: Eavesdropping or man in the middle attack on the wired control channel} 
\begin{itemize}
\item An attacker may be able to sniff the whole control traffic of multiple RRMUs which would allow him to get inside views of future behavior or configuration of the APs.
\item \textit{Countermeasure:} The communication over the control channel is encrypted by utilizing a 1-hop cryptography key. Every RRMU embeds its currently used symmetric group key within its probe request and response frames and uses this key for all outgoing traffic. Enhanced security between distinct peers is achieved by encrypting unicast messages using the public keys exchanged during the discovery phase.
\end{itemize}
\subsection{Thread: Rogue Attack} 
\begin{itemize}
\item A malicious user may be able to drive through an area and collect the credentials to build up the wired control channel to multiple local RRMUs which would allow him to influence their behavior in a malicious way.
\item \textit{Countermeasure:} ResFi RRMUs periodically
change the utilized group encryption session keys in
irregular time  intervals via the local wireless channel. The interceptor would have to place a local "spy device" remaining in a continuous connection with him. Nevertheless, on the other hand if he will try to act as his fixed IP address could be checked in any case of doubt (irregular or suspicious behavior). 
\end{itemize}

\subsection{Authenticity of the transmitting party}
\begin{itemize}
\item Again, even "drive-through" interception of the primary security material will not help. ResFi provides authenticity by the requirement that all outgoing ResFi messages sent via the wired backhaul have to be signed with the private key of the sender which allows the receiver to validate the signature with the corresponding public key exchanged during the discovery phase.
\end{itemize}

\subsection{Thread: Replay Attack} 
\begin{itemize}
\item If an attacker may be able to sniff control packets and send them unaltered but delayed to the original receivers this could result in confusion or misbehavior of the receiver APs RRMUs.
\item \textit{Countermeasure:} All sent ResFi messages are equipped with a unique
sequence number
\end{itemize}

\chapter{ResFi -- Detailed Specification}\label{sec:resfi_arch}

\section{Bootstrapping}\label{sec:boot}

\begin{figure}
 \centering
 \begin{minipage}[h!]{1\linewidth}
    \begin{center}
        \includegraphics[width=0.75\linewidth]{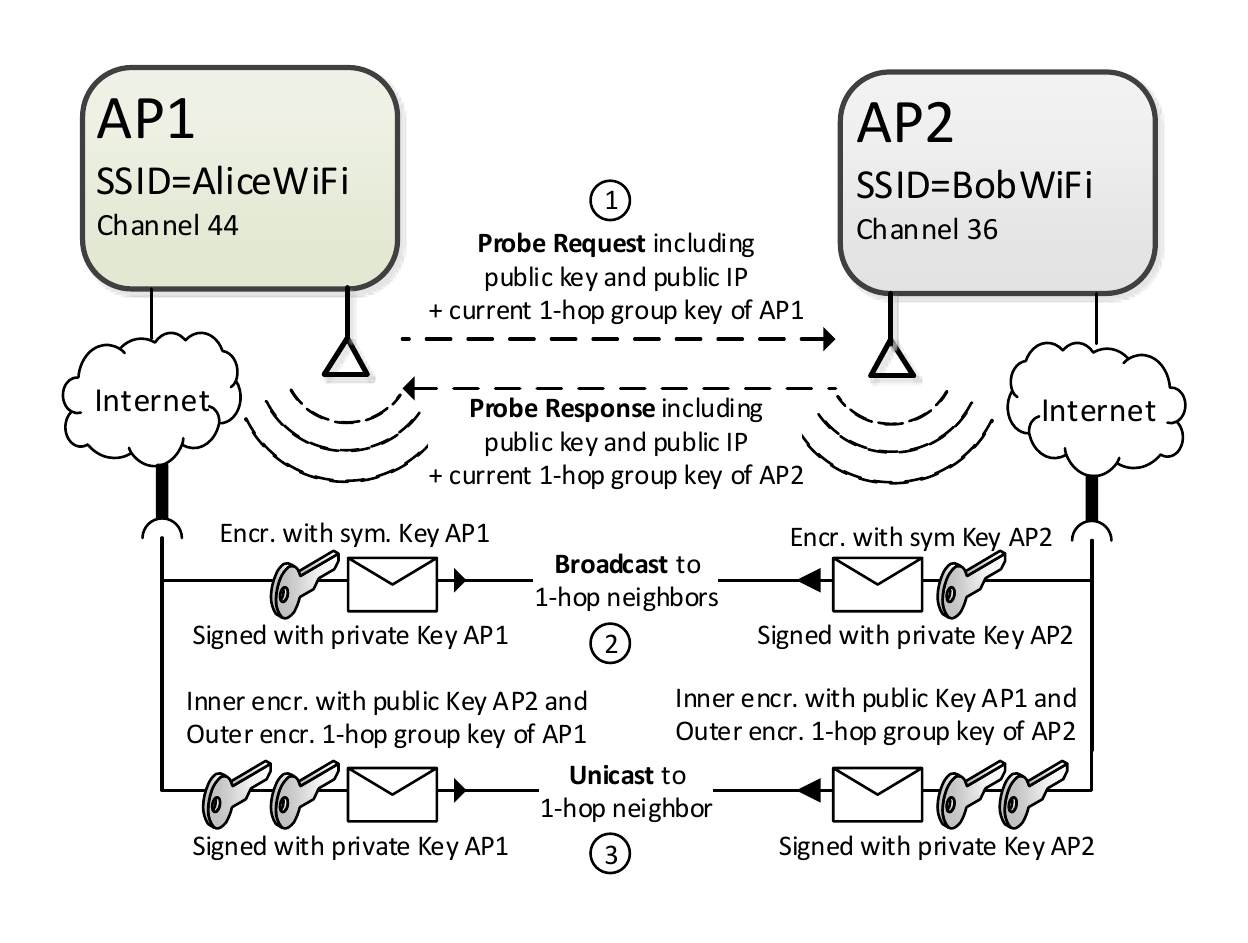}
    \end{center}
     \vspace{-15pt}
    \caption{Overview of the system architecture of ResFi: the wireless channel is used for exchange of configuration parameters (global IP of RRM unit, transient group encryption key and public RSA key) which are afterwards used for setting up the secure P2P out-of-band control channels over the Internet.}
    \label{fig:arch}
 \end{minipage}
 \vspace{-10pt}
 \end{figure}

After an ResFi enabled residential WiFi AP has booted up, the ResFi agent is started, the first symmetric group key and the RSA key pairs are generated and the discovery process is initiated. For each detected adjacent AP a mutual key and public IP exchange is performed over the wireless channel. This process is also depicted in Fig. \ref{fig:arch}.\\
In a first step the ResFi agent of the newly booted up AP ($AP_{0}$) performs a full active scan on all available IEEE 802.11 RF channels. This step includes the sending of a probe request\footnote{$\approx$ 212 octets standard probe response size or $\approx$ 64 octets standard probe request (depends on number of capabilities broadcasted in general by AP), plus each time the size of the vendor specific big ResFi IE (IE header 6 octets + transient group encryption key and IV 32 octets + 15 octets IP address + 162 octets DER encoded RSA public key)\label{fn:probesize}} including $AP_{0}$'s ResFi credentials (public IP of the RRM unit, currently used encryption key and public RSA key, embedded in an IE within the probe request) on each available RF channel which in turn triggers all ResFi APs in vicinity ($AP_{1..n}$) to send out their ResFi credentials embedded in an IE within a probe response\textsuperscript{\ref{fn:probesize}} back to $AP_{0}$, cf. Fig. \ref{fig:arch} tag 1. $AP_{0}$ subscribes itself to the publish (Pub) sockets of $AP_{1..n}$ using the public IP provided by the Probe Responses and $AP_{1..n}$ subscribe themselves to the publish (Pub) socket of $AP_{0}$ using the public IP provided by the Probe Request. Now $AP_{0}$ is able to successfully receive, validate and decrypt all messages sent via the wired backhaul by $AP_{1..n}$ and $AP_{1..n}$ are able to successfully receive, validate and decrypt all messages sent via the wired backhaul by $AP_{0}$.

The formation of the secure bidirectional control channel is completed. Broadcast messages to all neighbors are encrypted using the transient symmetric group key, cf. Fig. \ref{fig:arch}, tag 2 while unicast messages are in advance encrypted using the public RSA key of the corresponding receiver, cf. Fig. \ref{fig:arch}, tag 3. Moreover, all sent messages are signed using the private RSA key of the corresponding sender. This process is also described in Fig.~\ref{fig:fsm_1} as UML state machine.

\subsection{Standard Mode of Operation}

In the standard mode of operation the secure bidirectional control channel was already successfully established. All participating ResFi AP RRMUs are able to broadcast messages encrypted with their own group session key and signed with their own private RSA key to all of its one-hop ResFi neighbor RRMUs via the backhaul overlay network. All participating ResFi neighbor RRMUs are able to decrypt these messages, verify their integrity and the authenticity of the sender AP as a result of the mutual configuration data exchange in the bootstrapping phase. 
In addition to the standard operation of encrypting and signing outgoing messages and decrypting and verifying incoming messages, ResFi APs enable to broadcast messages to N-hop neighbors by performing TTL based forwarding.
Unicast messages which are in addition encrypted with the public RSA key of the corresponding receiver can only be sent within the one-hop neighbor group. If needed, multi hop unicast messaging with end-to-end encryption can be implemented on application level, cf. Sec. \ref{sec:cluster}.

\subsection{Transient Group Encryption Key}
During the standard mode of operation no specific control messages except the key change messages (KCM) have to be exchanged to enable the work of the distributed network. The object of the random periodic KCM and therefore of the group encryption key change is twofold, first it provides confidentiality on the wired backhaul channel and second it ensures that every group participant is a real physical neighbor located in wireless transmission range.\\ 
All ResFi agents have the obligation to periodically change their group encryption key and inform the other group members by sending a KCM as broadcast via the wired backhaul channel. The key change interval is bounded to $\text{KCMI}$ plus an randomly generated jitter. If a participating ResFi AP has not sent a KCM during $2 \cdot \text{KCMI}$ all other group members are removing the subscription to its publish socket.
A KCM always includes the current radio channel and the SSID of the sender to allow the other group members to use a single active frequency scan to obtain the new group session encryption key.
For the KCM always the old group key is used, while all messages sent after the KCM are encrypted using the new group key.
In advance the new group key is set within the probe response and probe request ResFi IEs for all new probe response and probe request messages.\\ ResFi APs that receive a KCM perform a single frequency active scan for the given RF channel and the given SSID which results in the reception of the new group key as described in the bootstrapping section. For the single frequency scans during runtime a empty probe request is used to trigger the KCM sender to reply with a probe response including the new group session key\footnote{small ResFi IE (IE header 6 octets + transient group encryption key and IV 32 octets)}. As ResFi relies on FIFO sockets and the scan procedure is blocking, all messages following the KCM, encrypted with the new group key, can always be decrypted successfully. Using the KCM scheme and single frequency / SSID scans performed by neighbor APs, the necessity of performing a new full active scan by the key changing AP is avoided. This prevents long deafness times due to active scanning on other RF channels.

\subsection{IP Address Change}

If the public IP address of a ResFi agent changes, the connectivity to all neighboring ResFi APs is broken. To overcome the connectivity loss, the affected ResFi agent repeats the bootstrap procedure described in Section \ref{sec:boot}.

\subsection{Radio Channel Change}
As the wireless channel after the boot-up phase is only used to obtain the symmetric group encryption key updates whose retrieval is always triggered by a KCM, which always includes the currently used radio channel, radio channel changing does not interfere the standard mode of operation of ResFi.

\begin{table*}[t]
\caption{ResFi north-bound API description}
\centering
\footnotesize
\begin{tabular}{ | p{.35\linewidth}p{.58\linewidth} | }
\hline
\rowcolor{BgGray2} \textbf{North-bound general framework API} & \textbf{Description}\\
\textit{getNeighbors()} & returns list of current neighbor IDs.\\
\textit{sendToNeighbor(nodeID, json\_msg)} & sends a JSON message to particular neighboring AP additionally encrypted using the public key of the receiver.\\
\textit{sendToNeighbors(json\_msg, TTL)} & sends JSON broadcast message to each direct neighboring AP, if TTL is used, flooding to N-Hop neighbors is performed.\\
\textit{regCallbacks(rxCb. newLinkCb, linkFailureCb)} & register callback functions used to deliver data to application (rxCallback $\rightarrow$ new message for application was received, newLinkCallback $\rightarrow$ a new neighbor detected, linkFailureCallback $\rightarrow$ neighbor was disconnected).\\
\textit{registerNewApplication(naming\_pattern)} & To handle parallel ResFi applications, name space separation for message handling is used.\\
\textit{getResFiCredentials(param)} & if \textit{param} == 1 returns public IP of RRMU, if \textit{param} == 2 returns public RSA key\\
\textit{usePrivateRSAKey(data, mode)} & enables to utilize the private key of the RRMU. If \textit{mode} == 1, returns signature computed over \textit{data}, if mode == 2, function decrypts \textit{data} and returns plaintext.\\
\rowcolor{BgGray2} \textbf{North-bound RRM API (suggestion)} & \textbf{Description}\\
\textit{getNetworkLoad(type)} & returns current network load: 1=number of served STAs, 2=total TX Bytes in DL, etc.\\
\textit{getChannels()} & returns available RF channels.\\
\textit{setChannel(chan)} & set (primary) RF channel to be used\\
\textit{setTxPower(mac\_addr, dbm)} & set transmit power towards STA with mac\_addr\\
\textit{setChannelWidth(mac\_addr, value)} & set channel BW for transmission to STA mac\_addr, e.g. 20, 40, 80, 160 MHz in 802.11ac\\
\textit{injectFrame(data)} & inject raw 802.11 frame\\
\textit{enableRTSCTS(mac\_addr, bool)} & enable usage of RTS/CTS towards STA with mac\_addr\\
\textit{startVAP(ssid, rxcb)} & start virtual AP with SSID, rxcb callback delivers received raw 802.11 frames.\\
\textit{deauthenticateSTA(mac)} & deauthenticate currently associated STA\\
\hline
\end{tabular}
\label{table:northboundapi}
\vspace{-10pt}
\end{table*}

\begin{table*}[t]
\caption{ResFi south-bound API description}
\centering
\footnotesize
\begin{tabular}{ | p{.35\linewidth}p{.58\linewidth} | }
\hline
\rowcolor{BgGray2} \textbf{South-bound framework API} & \textbf{Description}\\
\textit{getWiredInterface()} & enables ResFi to get the wired interface with IP access to backhaul Internet.\\
\textit{subscribeToProbeRequests()} & enables ResFi to retrieve the probe request payload from incoming probe requests.\\
\textit{addIEtoProbeResponses()} & enables ResFi to add/modify additional IE(s) to probe responses\\
\textit{performActiveScan()} & enables ResFi to start full/single active scan, takes add. IE which is added to probe req.\\
\rowcolor{BgGray2} \textbf{South-bound RRM API (suggestion)} & \textbf{Description}\\
\textit{\{set$|$get\}RfChannel()} & get/set currently used RF channel\\
\textit{\{set$|$get\}txPower(mac\_addr)} & get/set transmission power to be used to STA mac\_addr\\
\textit{\{set$|$get\}channelWidth(mac\_addr)} & get/set channel bandwidth to be used towards STA mac\_addr\\
\textit{\{set$|$get\}ClientInfo()} & get information about associated STAs (e.g. MAC, capabilities, RSSI, RX/TX count, rate statistics) or modify settings (e.g. set fixed rate, disconnect, priority, RTS/CTS usage, disassociate STA, associate STA, blacklist/whitelist STA)\\
\textit{\{getRx$|$getTx\}Stats(mac\_addr)} & get information about sent/received packets and bytes towards STA mac\_addr\\
\textit{injectRawFrame(data)} & inject raw 802.11 frame into wireless interface\\
\textit{startVAP(ssid, buffer)} & start new virtual AP with given SSID, all incoming data is saved in buffer.\\
\hline
\end{tabular}
\label{table:southboundapi}
\vspace{-10pt}
\end{table*}

\section{North-bound API}\label{sec:northboundapi}

The northbound algorithm/application API provided by ResFi is shown in Table~\ref{table:northboundapi}. The API is quite simple. Using the API any application is able to disseminate JSON messages to either APs in direct wireless communication range or to perform a general N-Hop TTL based flooding operation. Furthermore, unicast communication to direct peers is also available. If a new message via the framework is received the message processing can be controlled by registering a callback. ResFi determines the wireless context transparently for the user.

\section{South-bound API}
The ResFi framework can be easily integrated in existing AP solutions by connecting the existing platform to the ResFi southbound framework API listed in Table~\ref{table:southboundapi}. While the framework south-bound API is mandatory, the southbound API for the RRM is only a suggestion and can be extended to meet further application or algorithm needs. For this reason Table~\ref{table:southboundapi} only provides a subset of possible functions, in particular the RRM related part of the northbound API shows the required functions needed for the example applications in Section~\ref{sec:resfi_applications}.\\

\chapter{ResFi -- Implementation Details}\label{sec:bigap_implementation}

\begin{figure}
\centering
\begin{minipage}[h!]{0.75\linewidth}
	\begin{center}
			\includegraphics[width=0.6\linewidth]{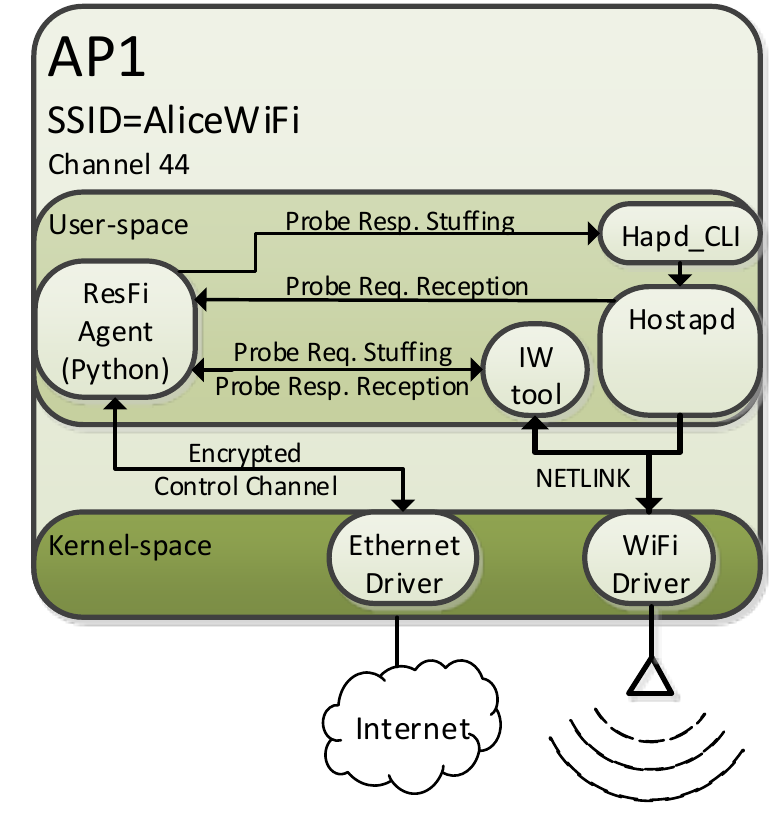}
	\end{center}
	 \vspace{-15pt}
	\caption{Overview of components in the ResFi prototype.}
	\label{fig:imp}
\end{minipage}
\vspace{-10pt}
\end{figure}

The ResFi implementation consists of the three components shown in Fig. \ref{fig:imp}. The ResFi framework agent is connected via the framework southbound API to a modified version of the software AP implementation \textit{Hostapd} \cite{hostapd-2013} via the also modified interface \textit{Hostapd\_CLI} which enables the embedding of additional IE(s) within probe responses, and the \textit{IW tool} which is used as an interface to trigger a new WiFi scan and to retrieve its results. Further the retrieval of the probe request payloads is realized using inter process communication (IPC) between hostapd and the ResFi agent. For our prototype we utilized standard x86 machines running Ubuntu 14.04 LTS. As the ResFi agent is programmed using platform independent Python code, it can be easily ported to various platforms. As the southbound API prototype realization is Linux specific it can be easily installed on all Linux based systems e.g. OpenWRT based APs or mobile AP solutions like Android smart-phones in tethering mode. 
The following sections describe the single prototype parts in more detail.

\section{ResFi Agent}

The ResFi Agent is implemented in Python and runs in user-space. The publish/subscribe (Pub/Sub) sockets for the back-haul wired overlay network are implemented using the Python \textit{\O{}MQ} library \cite{zeromq-2014}. On top of \O{}MQ the \textit{JavaScript Object Notation (JSON)} is used to serialize the data. Detection of IP address changes is implemented via a Netlink event callback. To provide authenticity, integrity and unicast confidentiality, 1024 bit \textit{RSA} key pairs are used and for group communication confidentiality, symmetric session encryption is performed using the \textit{Advanced Encryption Standard (AES)} in \textit{Cipher Feedback Mode (CFB)} with 128 bit key size. All security related functionality has been implemented by utilizing the \textit{PyCrypto Library} \cite{pycrypto-2015} and can be easily adapted to the needed purpose (e.g. different key-size, cipher mode or algorithm).

\section{Hostapd and Hostapd\_CLI}

\begin{figure}
\centering
\begin{minipage}[h!]{1\linewidth}
	\begin{center}
			\includegraphics[width=0.95\linewidth]{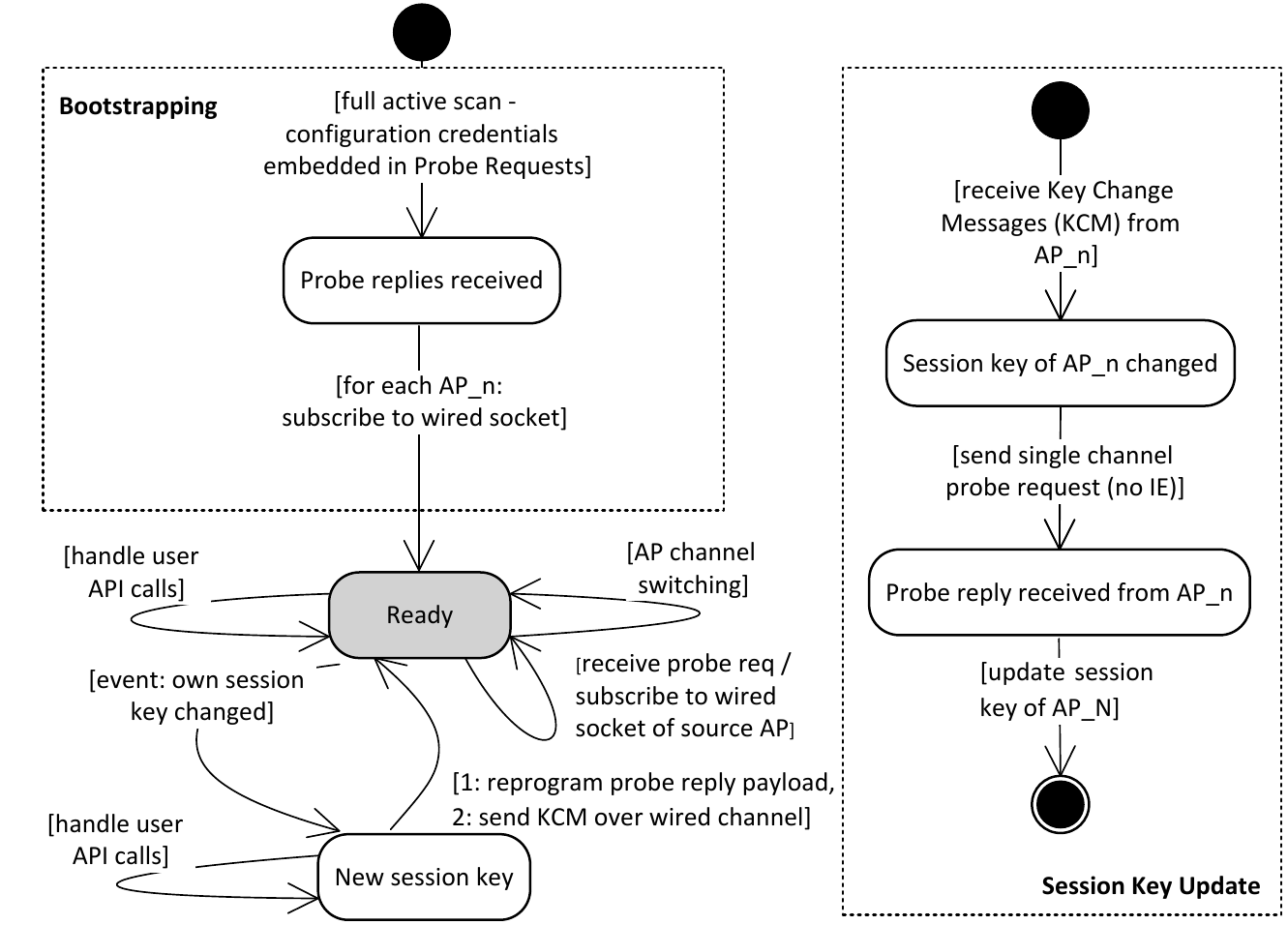}
	\end{center}
	 \vspace{-15pt}
	\caption{UML state machines describing the behaviour of the ResFi agent.}
	\label{fig:fsm_1}
\end{minipage}
\vspace{-10pt}
\end{figure}

Hostapd is responsible for performing all the AP management functionality on Linux based platforms. This includes the handling of probe requests and sending the probe responses. We modified hostapd in version 2.1 and the runtime interface hostapd\_cli to enable first, the embedding of additional IE(s) to all probe responses and second, the retrieval of the IE(s) from all received probe requests. Besides also the retrieval of the current AP parameters is enabled. The ResFi Agent calls hostapd\_cli to embed the public IP and the security keys into the probe response frames and to read the AP parameters. The probe request payload is retrieved using an additional \O{}MQ Pub/Sub socket to allow IPC between hostapd and the ResFi agent. The UML state diagram in Fig. \ref{fig:fsm_1} describes this inter-working in more detail.

\section{IW Tool}\label{sec:iw}
The Linux wireless WiFi configuration utility (IW) tool \cite{iw-2015} can be used to configure the WiFi driver in kernel-space from user-space. IW internally uses Netlink communication and the nl80211 library to enable user-space / kernel-space communication. We utilized the IW tool in version 4.3. The ResFi Agent calls IW to start and retrieve the results of a active WiFi scan on a single or over multiple channels and for a specific or unspecific SSID. Moreover, the IW tool is used to embed the additional ResFi IE(s) within the probe request messages used during the boot process.

\chapter{ResFi Mininet Emulation}

In order to offer the application developer an easy way to test own RRM algorithms, before deploying them in a real testbed, the ResFi framework allows the emulation of typical residential networks taking both the wireless access as well as the wired backbone network into account. This is achieved by running ResFi in Mininet~\cite{mininetUrl}, a container-based emulation which is able to emulate large network topologies on a single computer. Specifically, we emulate the wired Internet backhaul using the reported last-mile latencies and throughput values from~\cite{sundaresan2011broadband}. Moreover, the wireless channel which is used by ResFi for exchanging wireless management frames is also emulated. This is achieved using the following model: all APs in mutual wireless reception range are connected via a bidirectional link with fixed bandwidth (6 Mbps in case of 802.11g/a), delay (depending on distance) and loss characteristics (configurable parameter) to the same switch. Finally, the AP density which defines the wireless topology is a configurable parameter.

Note, any application code which was tested in the emulation environment can be used afterwards to be deployed on real hardware without any modifications. The ResFi Mininet Emulation is part of the ResFi framework which is provided as open-source under \url{https://github.com/resfi}.

\chapter{ResFi Application Examples}\label{sec:resfi_applications}

Next we present examples for applications supported by ResFi as a showcase.

\section{Network Clustering}\label{sec:cluster}

In order to reduce the information update overhead and to optimize the use of the network bandwidth obtaining a hierarchical organization of the residential AP network is desired. This can be achived by clustering algorithms that partition the AP nodes of network into clusters~\cite{basagni1999distributed}. A clustering is crucial for controlling the spatial reuse of the shared wireless channel (e.g., in terms of time division or frequency division schemes).

As a proof of concept we implemented both the Distributed Clustering Algorithm (DCA) and the Distributed Mobility-Adaptive Clustering (DMAC) proposed by Basagni~\cite{basagni1999distributed}
as applications in ResFi.
\section{End-to-End Security for N-Hop Neighbors}\label{sec:end-to-end-sec}
Basically, ResFi provides one-hop encryption by utilizing a group encryption key between all one-hop neighbors. In addition unicast messages to one-hop neighbors are encrypted using the public key of the receiver. For N-hop messages, all intermediate hops decrypt the forwarding message locally with the last hop group encryption key and forward it encrypted with their own group encryption key. If end-to-end security is needed between N-hop neighbors, this functionality can be easily implemented as ResFi application. E.g. to enable encrypted communication in a established cluster, cf. Sec. \ref{sec:cluster}, the cluster head can utilize the \textit{getResFiCredentials()} function to obtain its public key and propagate it to all cluster nodes and vice versa. As now all participants know the public key of each other end-to-end signing and en/decrypting of messages is possible using the function \textit{usePrivateRSAkey()}, see Table~\ref{table:northboundapi}.
\section{Dynamic Channel Selection}
\label{sec:chansec}

The ResFi framework allows an easy implementation of distributed dynamic channel selection schemes for WiFi APs. According to the approach proposed by Mishra et al.~\cite{mishra2005weighted} each AP may periodically inform its direct neighbor APs about its network load (e.g., number of served clients or flows), recent airtime utilization on different channels, the presence of WiFi and non-WiFi networks and its own radio channel. Such information can be combined at each AP to select the least congested channel. As a proof-of-concept we implemented the aforementioned algorithm (Lst.~\ref{dist_chan_algo}).

\lstset{language=Python, 
        basicstyle=\ttfamily\scriptsize, 
        keywordstyle=\color{deepblue},
        commentstyle=\color{comments},
        stringstyle=\color{deepgreen},
				emphstyle=\ttb\color{deepred},    %
        showstringspaces=false,
        procnamekeys={def,class}}
 
\begin{lstlisting}[caption=Distributed channel assignment implemented using ResFi., label=dist_chan_algo]
def channelSelection():
  agent = ResFiAgent() # init ResFi agent
  agent.regCallbacks(rx_cb, Null, Null)
  agent.registerNewApplication(de.tu-berlin.ch-assign)
  apNode = agent.getResFiCredentials(1) # get node ID (IP)
  channel = 0 # init with channel 0
  load = getNetworkLoad() # get network load
  chs = getChannels() # number of radio channels available
  nbInfo = {} # neighbor info

  while True:
    msg = {'node': apNode, 'ch': channel, 'load': load}
    agent.sendToNeighbors(msg) # API call
    time.sleep(random.uniform(0, jitter/2)) # backoff

def rx_cb(json_msg): # receive callback function
  sender = json_msg['node'], nb_channel = json_msg['ch']
  nb_load = json_msg['load']
  nbInfo[sender] = {'load': nb_load, 'ch': nb_channel}

  # calc Hc according to Hminmax algorithm:
  Hc = {}
  for c in range(chs): # for each channel
    Hc[c] = 0 # reset to zero
    for entry in nbInfo: # for each neighbor
      tmpCh = nbInfo[entry]['ch']
      if tmpCh == c: # same channel
        # select the max() weight; here load
        Hc[c] = max(Hc[c], load + nbInfo[entry]['load'])

  # choose channel with minimum Hc
  channel = getChannelWithMinConflictWeight(Hc)
  agent.setChannel(channel)
\end{lstlisting}

\section{Interference Management}
The well-known hidden terminal problem~\cite{shrivastava2009centaur} causes severe co-channel interference (and thus packet loss) in dense WiFi networks with multiple APs operating on the same radio channel. While the use of virtual channel reservation has a potential to reduce the number of hidden nodes it creates significant overhead by exchange of IEEE 802.11 RTS/CTS packets. Therefore, an adaptive RTS/CTS scheme activated only on wirleess links suffering from hidden terminal problem would be favorable. This can be easily achieved using our ResFi platform. For this purpose each AP could perform passive hidden terminal detection as proposed in~\cite{li2006passive} and inform its neighboring APs about links potentially affected by hidden terminals for which the RTC/CTS handshake would be enabled, see function \textit{enableRTSCTS()} in Table~\ref{table:northboundapi}.

\section{Virtual Access Points (VAP)}\label{sec:vap}

The spatial area covered by a single WiFi AP is limited especially when using the 5\,GHz ISM band with unfavorable propagation characteristics. In dense residential areas there is a high probability that a significant parts of a residential apartment is in excellent coverage of neighbor's AP rather than within the range of its own home AP~\cite{Shi-2015}. A way to utilize the neighboring AP is to deploy on-demand a virtual AP on the neighboring AP and to tunnel all encrypted WiFi traffic to the home AP~\cite{vestin2013cloudmac}. This allows the client devices to always authenticate against the home AP using the WPA passphrase already stored in the device. There is no registration process; no software to install on the device; not even any settings to change.

The on-demand deployment of VAPs can be easily achieved using the ResFi framework. Specifically, each AP has to disseminate information about the configured SSIDs in its home AP to the neighboring AP where dynamically a VAP is configured, see function \textit{startVAP()} in Table~\ref{table:northboundapi}.

\section{Client STA Handover for Load Balancing and Mobility Support}
The BIGAP approach~\cite{Zubow15bigap_seamless_handover} which enables soft handover operations in centralized enterprise WiFi networks, can also be implemented as ResFi application. If combined with the VAP application (cf. Sec.~\ref{sec:vap}), soft handover between the home AP and the neighboring AP to enable mobility and load-balancing support without network outage, can be realized. If the client STA supports dynamic frequency selection (DFS) and both, the current AP and the target AP are operating on different RF channels, soft handover operations are possible by injecting an additional beacon frame including a channel switch announcement IE with the RF channel of the target AP via the function \textit{injectFrame()} executed on the current AP. If no DFS support on the client STA is available, hard-handover using the function \textit{deauthenticateSTA()} on current AP enables a controlled handover. All aforementioned functions are part of ResFi's NB API, see Table~\ref{table:northboundapi}.
\chapter{Evaluation}\label{sec:evaluation}

Our proposed framework is analyzed and evaluated by means of experiments in a WiFi testbed. Moreover, the expected control overhead in the wireless channel is analyzed analytically. 
An overview in which the overhead generated by the ResFi framework in its different mode of operation caused by channel switching and sending messages either via the wired backhaul or via the wireless channel is given in Table \ref{table:overhead}.
The evaluation results of our proposed ResFi approach are further compared with the approach proposed by Manweiler~\cite{manweiler-2012} extended to multi-channel environments to which refer as RxIP++.

\section{Active vs. passive Scanning}\label{sec:scan}
As described in Section \ref{sec:resfi_arch}, ResFi uses different modes of scanning depending on its operation state. In the boot-up phase in which the deafness of the AP is not as important as during the standard mode of operation, a full scan is used. After the boot-up is completed, a single frequency scan for a single SSID is utilized to keep the AP deafness as short as possible. To enable the analysis of the overhead of this approach we measured the timing of different active and passive scanning procedures (cf. Sec. \ref{sec:dot11primer}) using diverse WiFi hardware. The results of this experiment are later used in Sec.~\ref{sec:reclat} to calculate the overhead due to our frequent symmetric group encryption key change via the wireless channel.
\subsection{Methodology}
In this experiment we used different commercial off-the-shelf (COTS) WiFi hardware two connected via PCI and two connected via USB. The utilized WiFi chips are listed in Fig.~\ref{fig:eval_scanning_times_full} and~\ref{fig:eval_scanning_times_single}. The experiments were executed on x86 machines with Ubuntu 14.04 LTS. The scanning calls were started using the command-line tool iw (cf. Sec. \ref{sec:iw}).

\subsection{Results}
\begin{figure}
 \centering
 \begin{minipage}[b]{1\linewidth}
    \begin{center}
        \includegraphics[width=0.7\linewidth]{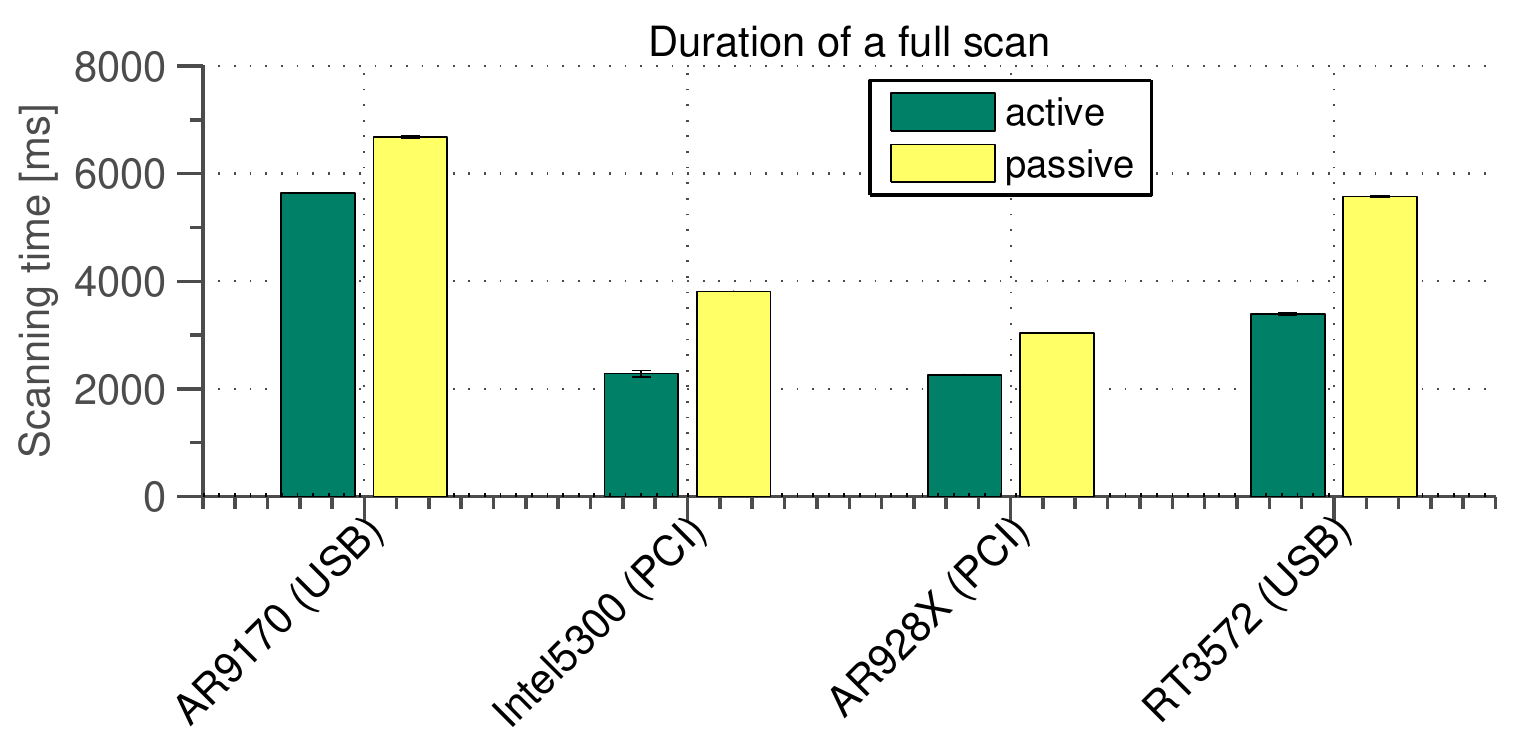} %
    \end{center}
     \vspace{-15pt}
    \caption{Scanning duration of a full scan (performed over all available WiFi channels, errorbar shows the standard deviation).}
    \label{fig:eval_scanning_times_full}
 \end{minipage}
 \vspace{-10pt}
 \end{figure}

The timing results of full WiFi scans are shown in Fig.~\ref{fig:eval_scanning_times_full}. We see that independent of the scanning mode, the scanning durations strongly vary between different WiFi chips and connection technology. in general, chipsets connected via PCI show shorter scanning durations while USB connected chips are slower. However, when the two different scanning modes (passive or active) are evaluated it is obvious that active scanning is always superior to passive scanning w.r.t. to the scanning latency.

\begin{figure}
 \centering
 \begin{minipage}[b]{1\linewidth}
    \begin{center}
        \includegraphics[width=0.7\linewidth]{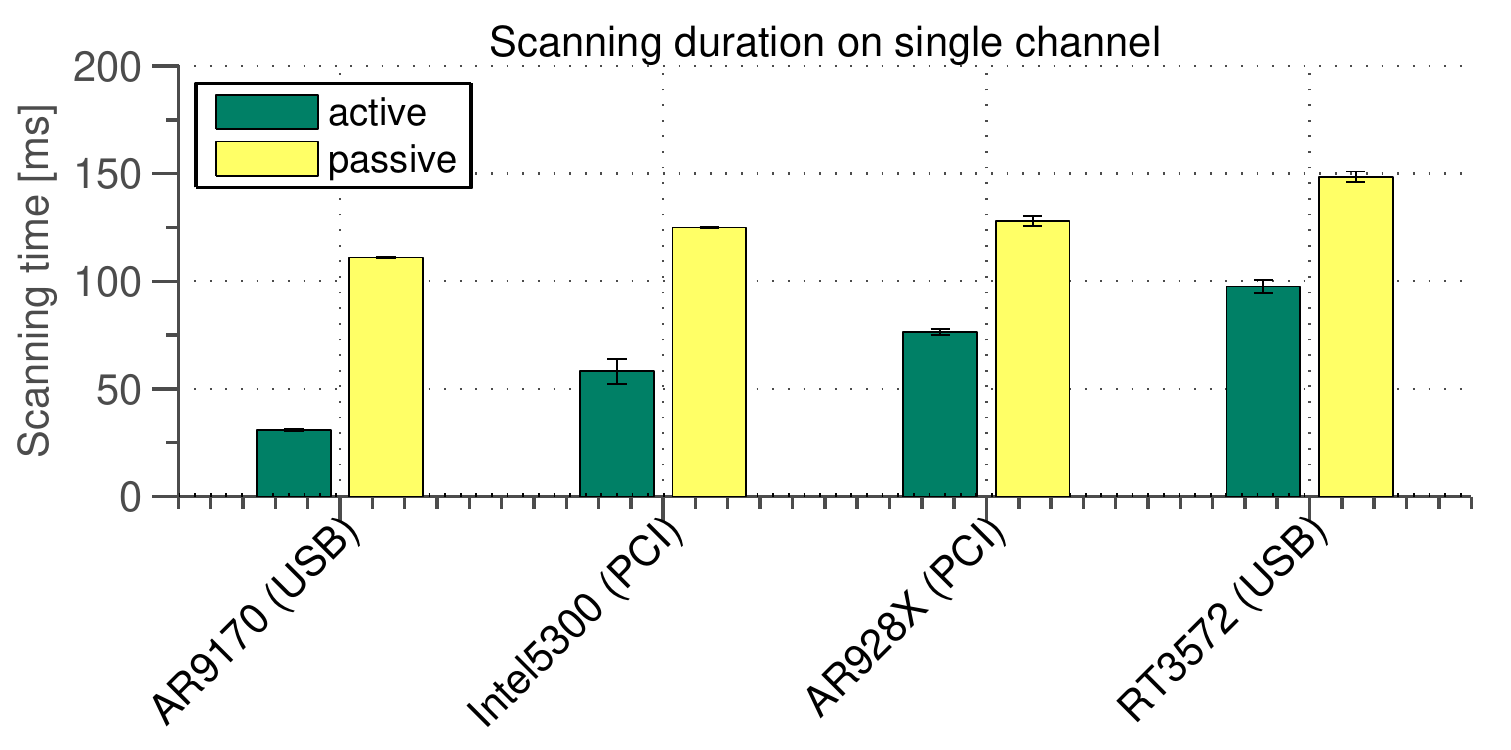} %
    \end{center}
     \vspace{-15pt}
    \caption{Scanning duration on single channel (errorbar shows the standard deviation).}
    \label{fig:eval_scanning_times_single}
 \end{minipage}
 \vspace{-10pt}
 \end{figure}

The results of the latency experiment of single frequency scans is depicted in Fig.~\ref{fig:eval_scanning_times_single}, interestingly the connection type whether USB or PCI does not effect the scanning latency. Nevertheless, for single frequency scans active scanning is also always faster than passive scanning.

\textbf{Takeaways:} Single frequency scans always provide the shortest latency in comparison to full scans (e.g. 30ms vs. 6300ms for AR9170). Active scanning is always faster than passive scanning.
 
\section{Reconfiguration Overhead}\label{sec:reclat}
In the following we are analyzing the overhead in the wireless channel due to frequently changing the transient symmetric group encryption key.

\subsection{Methodology}

The overhead in the wireless channel is due to the transmission of probe request and reply packets which are sent on a basic bitrate (e.g. 6\,Mbps in 802.11a/g). Moreover, during a scanning operation for neighboring nodes the AP is deaf and cannot handle data transmissions of its associated client stations and hence is wasting valuable airtime. Moreover, an associated station may disassociate if it misses too many beacon frames.
As shown in Section \ref{sec:scan}, the duration of a single active WiFi scan for a given SSID on a particular radio channel takes depending on the hardware between $30\,$ms and $100\,$ms.    
Hence, there is a tradeoff between the rate at which the reconfiguration takes place and the available airtime in the wireless channel for data communication.

Because in ResFi a reconfiguration at a single AP triggers the scanning operation of each neighboring AP the expected AP density plays a major role. For our analysis we analyzed the data provided bv the large-scale measurement campaign of Biswas et al.~\cite{biswas2015large, merakii-2015} whereas the number of neighboring APs in the 2.4 and 5\,GHz band is on average 16.8 and 5.1 respectively. To pay attention to virtual WiFi networks in which one physical AP broadcasts multiple SSID and BSSIDs, we only included BSSIDs into the results in which the RSSI, the OUI and the WiFi capabilities are different. Note, the overhead in 2.4\,Ghz is also larger because the management frames (here probe requests and replies) are sent on a lower PHY bitrate, i.e. 1 vs. 6\,Mbps.

\begin{table*}[t]
\caption{Overhead analysis.}
\centering
\footnotesize
\begin{tabular}{ |p{.11\linewidth}p{.17\linewidth}p{.30\linewidth}p{.30\linewidth}| }
\hline
\rowcolor{BgGray2} \multicolumn{2}{|c}{\textbf{Operation}} & \multicolumn{1}{c}{Radio (IEEE 802.11)} & \multicolumn{1}{c|}{Backhaul}\\
\hline
\multirow{6}{*}{Standard mode} & \multirow{2}{*}{\parbox{2.8cm}{change own group key}}& \multirow{2}{*}{\parbox{4.4cm}{for-each neighboring AP: send probe response + small ResFi IE}} & \multirow{2}{*}{\parbox{4.4cm}{for-each neighboring AP: send Key Change Message (KCM)}} \\
 & & & \\
 
 & \multirow{2}{*}{rec. KCM}& \multirow{2}{*}{\parbox{4.4cm}{channel switch \& send probe req. (no ResFi IE)}} & \multirow{2}{*}{\parbox{4.4cm}{zero}} \\
 & & & \\ 
 
  & \multirow{2}{*}{\parbox{2.8cm}{rec. probe req. + big ResFi IE}}& \multirow{2}{*}{send probe response + big ResFi IE} & \multirow{2}{*}{\parbox{5.4cm}{zero}} \\
  & & & \\
  
 & \multirow{2}{*}{\parbox{2.8cm}{rec. probe req. (no ResFi IE)}}& \multirow{2}{*}{send probe response + small ResFi IE} & \multirow{2}{*}{\parbox{5.4cm}{zero}} \\
    & & & \\
 
 \hline

\multirow{2}{*}{Bootstrapping} & & for-each available Rf channel: channel switch \& send probe request + big ResFi IE & \multirow{3}{*}{\parbox{6cm}{zero}} \\
 \hline
\end{tabular}
\label{table:overhead}
\vspace{-10pt}
\end{table*}

Next, we give a detailed description of the overhead analysis for both the beacon stuffing approach used in RxIP++ which serves as baseline and the ResFi approach using probe request and reply management frames.

\subsubsection{RxIP++}

When using the approach from RxIP for dissemination of configuration data the overhead is due to the transmission of additional IEs in the beacon frames and the required scanning overhead in multi-channel environments. The overhead depends on the AP density as each AP performs a reconfiguration, i.e. the larger it is the more wireless frames need to be exchanged and more scanning operations need to be performed. For a network of $N$ co-located, i.e. in communication range, APs the relative overhead for each AP can be computed as follows:
\begin{align}\label{eq:beacon_eq}
O_{\mathrm{RxIP++}} &= \frac{1}{C} \times N \times T_{\mathrm{Beacon-IE}} \times R_{\mathrm{Beacon}} + (N-1) T_{\mathrm{scan}}
\end{align}
\noindent where $C$ is the total number of channels available, $N$ is the number of neighboring APs, $T_{\mathrm{Beacon-IE}}$  and $R_{\mathrm{Beacon}}$ are the additional beacon overhead and beacon interval (10\,Hz) respectively. The first term represents the overhead due to the additional transmission of IE in beacon frames. Note, that due to multi-channel environment the APs are operating on different radio channels, hence to get the per channel overhead we have to divide by the number of channels. The second term represents the overhead due to scanning deafness.

\subsubsection{ResFi}

Next, we analyze the overhead of the ResFi approach:
\begin{align}\label{eq:proposed_eq}
O_{\mathrm{ResFi}} &= (N-1) (T_{\mathrm{PReq}} + T_{\mathrm{PRep}}) + (N-1) T_{\mathrm{scan}} \nonumber \\
& + \frac{1}{C} \times (N-1) \times (N-2) \times (T_{\mathrm{PReq}} + T_{\mathrm{PRep}})
\end{align}
\noindent where the first and third term represent the overhead due to transmission of probe request and reply messages and the second term accounts for deafness due to scanning procedure.

\subsection{Results}

 \begin{figure}
 \centering
 \begin{minipage}[b]{0.85\linewidth}
    \begin{center}
        \includegraphics[width=0.9\linewidth]{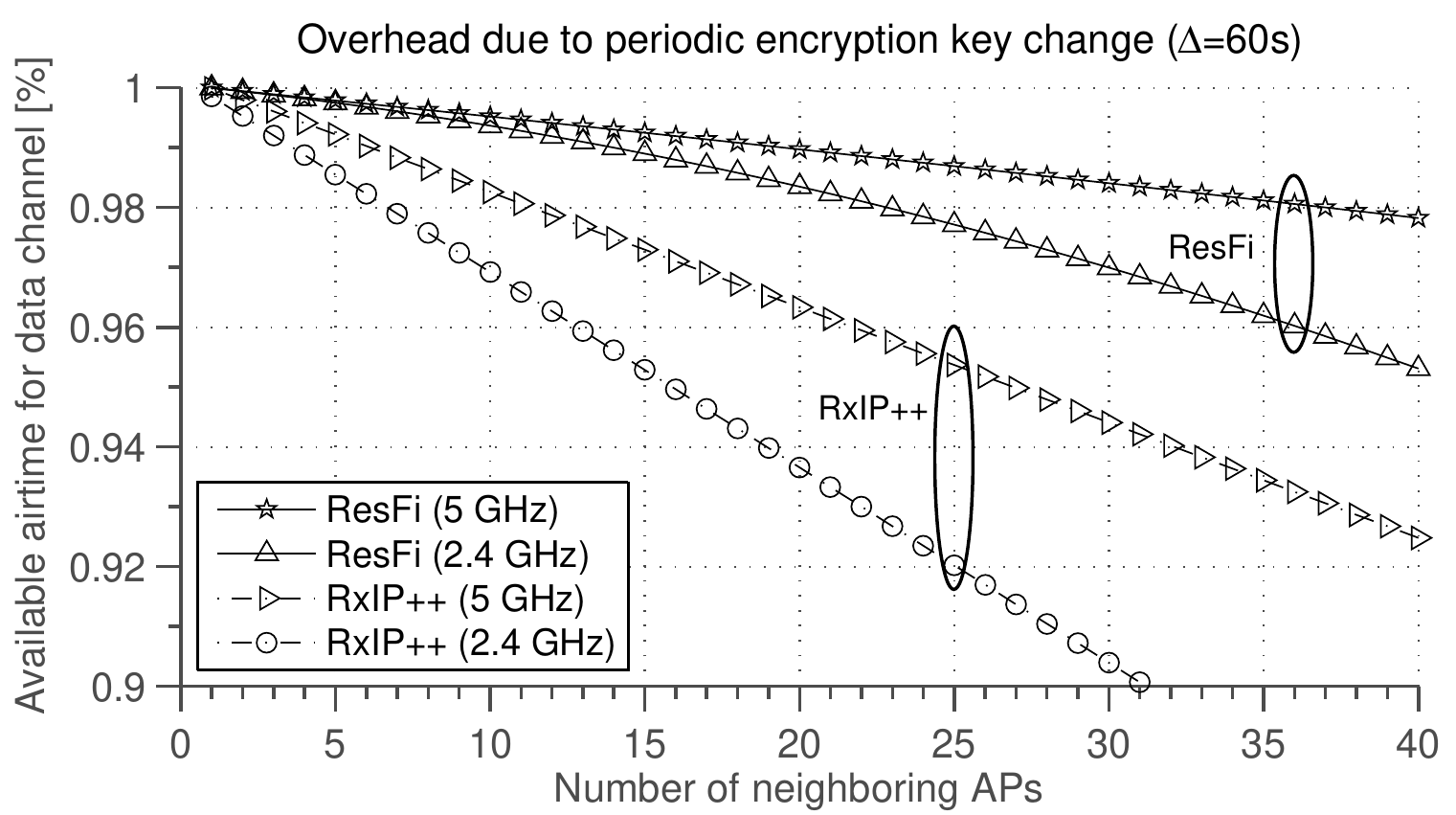}
    \end{center}
     \vspace{-15pt}
    \caption{Impact of periodic group encryption key change ($\Delta=60\,s$) on the available airtime in the data channel.}
    \label{fig:overhead_analytical}
 \end{minipage}
 \vspace{-10pt}
 \end{figure}

Using equations \ref{eq:beacon_eq} and \ref{eq:proposed_eq} we are able to calculate the overhead for different AP densities, i.e. number of neighboring APs. Here we assume that each AP performs a single group encryption key update. Fig.~\ref{fig:overhead_analytical} shows the relative available airtime in the wireless data channel with a update interval of 60s, i.e. $1-O_{\mathrm{proposed}}$ and $1-O_{\mathrm{Beacon}}$ respectively.

The results can be summarized as follows. In the 2.4 Ghz and the 5 Ghz band the overhead for a single reconfiguration is highest with RxIP++, whereas using the proposed ResFi approach which relies on probe request and probe response frames is superior in both bands for a reconfiguration period of $60\,$s and AP densities between 0 and 40.

 \begin{figure}
 \centering
 \begin{minipage}[b]{0.8\linewidth}
    \begin{center}
        \includegraphics[width=0.95\linewidth]{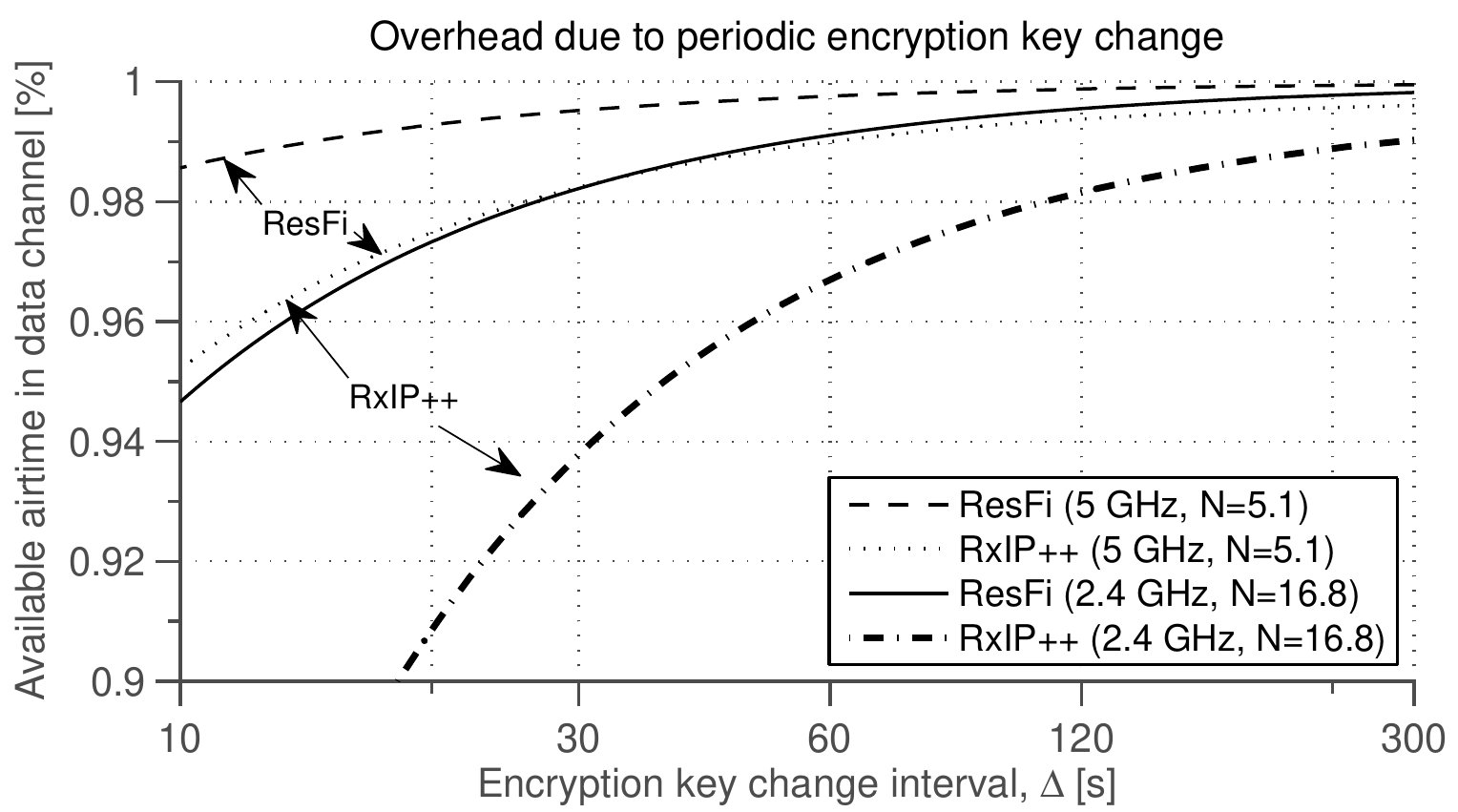} %
    \end{center}
     \vspace{-15pt}
    \caption{Tradeoff between encryption key change interval and available airtime in the data channel ($N$ represents the number of neighbors each AP has).}
    \label{fig:key_change_rate}
 \end{minipage}
 \vspace{-10pt}
 \end{figure}

Next, we analyze the impact of the reconfiguration rate on the available airtime in the wireless data channel. The results are shown in Fig.~\ref{fig:key_change_rate}. From a practical point of view a maximum overhead of 1\% is tolerable. Hence the maximum reconfiguration rate is pretty low, i.e. update every 60 and 20\,s for the 2.4 and the 5\,GHz band respectively. However, for the envisioned residential AP szenario it is still sufficient as we suggest to change the group encryption key every minute. Again the proposed ResFi approach is superior in both bands.

\textbf{Takeaways:} There is a clear tradeoff between reconfiguration rate and overhead in the wireless channel. The beacon-stuffing approach (RxIP++) is not feasible in real residential deployments with high AP densities.

\section{Reconfiguration Latency}

\subsection{Methodology}

In this experiment we analyze reconfiguration latency in ResFi due to changing configuration data, e.g. group encryption session key. The reconfiguration latency is composed of the delay due to transmission of the key change message (KCM) over the wired out-of-band control channel as well as the scanning delay due to active scanning on a particular channel and given SSID.

We considered two different wired backhaul technologies. First, Gigabit Ethernet as a very low latency backhaul which we use in our testbed. It serves as a baseline. Second, the typically used backhaul technology in residential WiFi deployment, i.e. cable/DSL. For the latter we used the traffic control tool~\cite{tcUrl} to emulate the last-mile latency in residential WiFi deployments as reported by~\cite{sundaresan2011broadband}. Note, the last-mile latency is the latency to the first hop inside the ISP’s network and hence captures the latency of the access link (DSL/cable). According to~\cite{sundaresan2011broadband} most users of cable ISPs are in the 0–10 ms interval whereas a significant proportion of DSL users have last-mile latencies of more than 20 ms, with some users seeing last-mile latencies up to 60 ms.

For the experiments we used x86 machines with Ubuntu 14.04 LTS and Linksys AE1000 WiFi USB sticks using Ralink rt2800 chipsets as APs.

\subsection{Results}\label{sec-scanning}

\begin{figure}
 \centering
 \begin{minipage}[b]{0.8\linewidth}
    \begin{center}
        \includegraphics[width=0.8\linewidth]{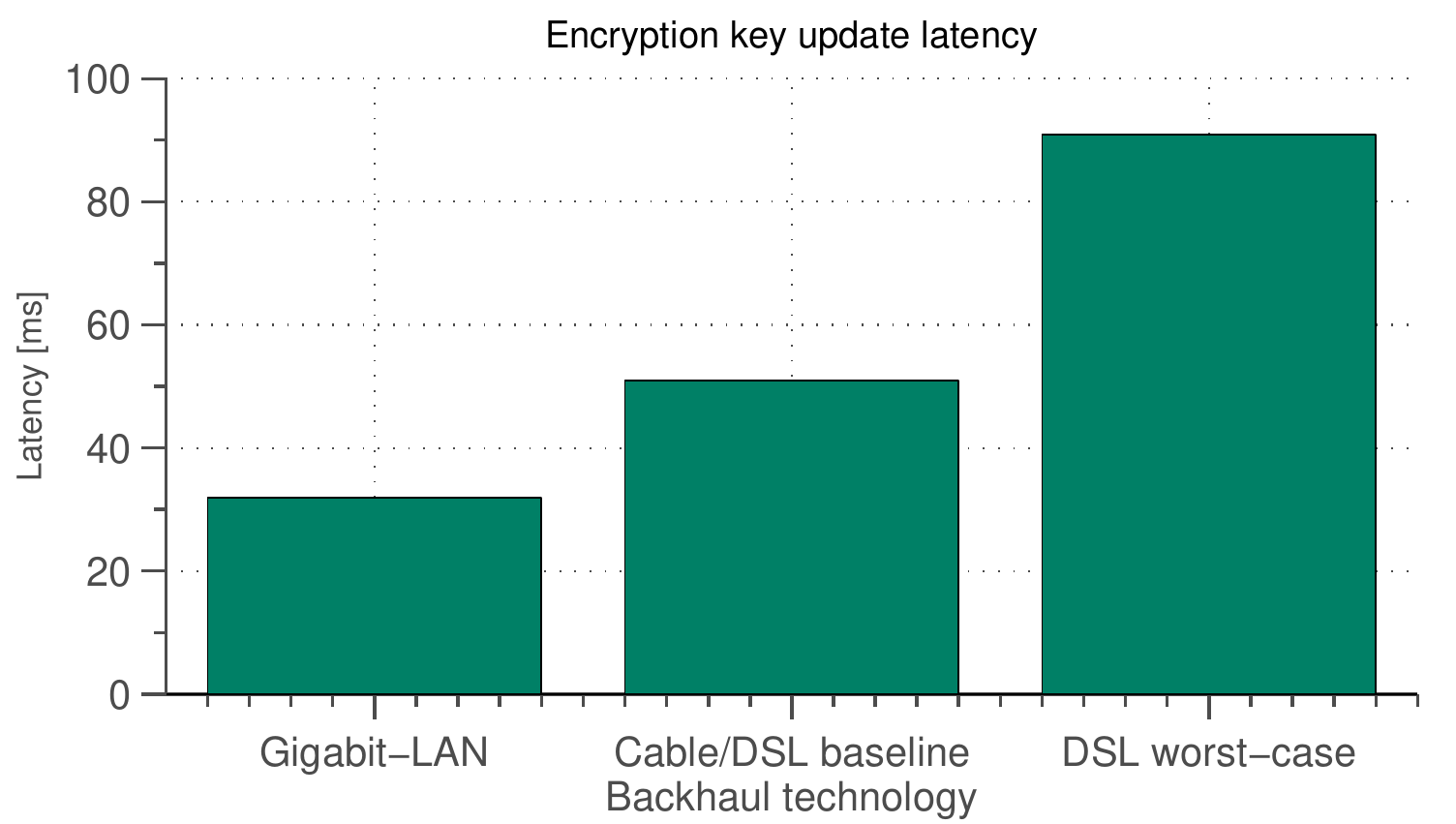} %
    \end{center}
     \vspace{-15pt}
    \caption{Reconfiguration latency due to changing encryption key (Confidence $\geq95\%$).}
    \label{fig:reconfig_latency}
 \end{minipage}
 \vspace{-10pt}
 \end{figure}

The results are shown in Fig.~\ref{fig:reconfig_latency}. We see that in the worst case, i.e. DSL, the reconfiguration latency is around 165\,ms which is 58\,\% higher as compared to Gigabit LAN. 

\textbf{Takeaways:} Due to the reconfiguration latency configuration data like the used group encryption session key exchanged on the wireless channel can be changed at most $6 \times$ per second when assuming the scenario with worst-case DSL backhaul access. This is more than sufficient to achieve a high level of security.

\chapter{Conclusions and Areas for further research}\label{sec:conclusions}

Up to our knowledge we have presented the first proposal of a holistic platform supporting automatic establishment of secure connectivity within a definable scope of neighborhood and set of resource management supporting functions for residential WiFi networks. Our proposal allows usage of legacy hardware and avoids violation of the existing management borders following out of the fragmented ownership structure. ResFi was prototypically implemented and the source code is provided to the community as open source. We believe that there is a clear need for such a solution. 

On our side the following further areas of work on this framework have been already identified:
\begin{enumerate}
\item Many RRM require rather tight time synchronization among the nodes. So far ResFi relies on Network Time Protocol (NTP~\cite{mills2010network}) to time synchronize over the Internet backhaul which achieves only an accuracy of 10s of ms in WAN networks. We intend to extend ResFi to provide over-the-air time synchronization using either 802.11 beacons~\cite{manweiler-2012} or using 802.11 management frames for exchanging IEEE 1588 Precision Time Protocol (PTP) frames~\cite{lee2005ieee}.

\item The semantics of the network load - a notion introduced in our API - is not unique. Different function of the air time utilization, number of neighbors etc. have been used in the past in this context. 
While in the actual version we consider the air time utilization on the actually used channel as the metric of the network load, we consider offering a possibility to introduce in a flexible way a definition of this parameter. 

\item The notion of one-hop neighborhood is not unique, either. At this moment we include in the one-hop neighborhood any AP which provides a decodable probe response to a probe request broadcasted with the lowest bit rate. This notion might be generalized by attributing to the probe exchange some constraints on power with which this exchange is performed.  
 
\end{enumerate}

While we believe to have covered a reasonable set of requirements while keeping the solution relatively simple, we have so far verified its merits only using a few very simple case studied. We hope that usage of this framework (enhanced by the open source approach) for more complex RRM functions might lead to its further improvement.

\bibliography{biblio}
\end{document}